\documentclass[12pt]{article}

\usepackage{a4wide}
\usepackage[final]{graphicx}
\usepackage{amsmath,amssymb}
\usepackage{authblk} 
\usepackage[numbers,compress]{natbib}
\usepackage[margin=1em,font={small},labelfont={sc}]{caption}

\usepackage[pagebackref=false,colorlinks=true,citecolor=blue,filecolor=blue,%
linkcolor=blue,anchorcolor=blue,urlcolor=blue,hypertex]{hyperref}

\renewcommand\Re{\operatorname{\mathfrak{Re}}}
\renewcommand\Im{\operatorname{\mathfrak{Im}}}

  \renewenvironment{thebibliography}[1]{%
    \begin{oldthebibliography}{#1}%
      \setlength{\parskip}{0ex}%
      \setlength{\itemsep}{0ex}%
      \small
  }%
  {%
    \end{oldthebibliography}%
  }

\begin{document}

\title{\LARGE \bf
Neural network generated parametrizations of deeply virtual
Compton form factors
}

\author[1]{Kre\v{s}imir Kumeri\v{c}ki}
\author[2]{Dieter M\"{u}ller}
\author[3]{Andreas Sch\"{a}fer}
\affil[1]{Department of Physics, University of Zagreb, Zagreb, Croatia}
\affil[2]{Lawrence Berkeley National Lab, Berkeley, USA}
\affil[2]{Institut f\"ur Theoretische Physik II, Ruhr-Universit\"at Bochum,  Germany}
\affil[3]{Institut f\"{u}r Theoretische Physik, Universit\"{a}t Regensburg, Regensburg, Germany}
\date{}
\maketitle

\begin{abstract}
\noindent
We have generated a parametrization of the Compton form factor (CFF)
$\mathcal{H}$ based on data from deeply virtual Compton scattering (DVCS)
using neural networks. This approach
offers an essentially model-independent fitting
procedure, which provides realistic uncertainties. Furthermore, it
facilitates propagation of uncertainties from
experimental data to CFFs.
We assumed dominance of the CFF $\mathcal{H}$ and used
HERMES data on DVCS off unpolarized protons.  We predict the
beam charge-spin asymmetry for a proton at the kinematics of the
COMPASS II experiment.
\end{abstract}

\section{Introduction}

Generalized parton distributions (GPDs)
\cite{Mueller:1998fv,Radyushkin:1996nd,Ji:1996nm}
provide a detailed description of the nucleon in terms of partonic degrees of
freedom, which allows, e.g., to address
the three-dimensional distribution of quarks and gluons and the partonic decomposition
of the nucleon spin  \cite{Ji:1996ek}. Their determination improves our understanding of
non-perturbative QCD dynamics in general. More specifically, they are a
necessary (though not sufficient) input
for the theoretical description of multiple-hard reactions in
proton-proton collisions at LHC collider \cite{Frankfurt:2003td,Blok:2010ge,Diehl:2011tt}
and for other applications reaching beyond the limit of collinear QCD.
Information on GPDs comes from many sides.
There exists already an impressive experimental data base for exclusive reactions, which is
continuously improved.
A standard approach is either confrontation of model
predictions \cite{Goloskokov:2007nt} or
least-squares fit \cite{Kumericki:2009uq} to this data.
The latter approach, formulated in Mellin space, is analogous to
the well-established global fitting framework of  parton distribution functions
(PDFs). Complementary information on moments of GPDs comes from lattice QCD
\cite{Bratt:2010jn}.

However, compared to the situation for PDFs, extracting GPDs from data is a much more
intricate task
and model ambiguities are much larger.
For instance, the  theoretically cleanest process, used for the determination of GPDs, is
deeply virtual Compton scattering (DVCS), $\gamma^{*} p \to \gamma p$,
which can be parametrized by twelve Compton form factors (CFFs). At leading order in the
(inverse) photon virtuality squared
$1/\mathcal{Q}^2$ they are expressed as convolutions of GPDs with perturbatively
calculable coefficient functions. At leading order in the QCD
coupling constant and for one of the CFFs, $\mathcal{H}$, we have
\begin{equation}
{\cal H} (x_B,t,{\cal Q}^2) \stackrel{\rm LO}{=}
  \int_{-1}^1\!dx\, \left[\frac{1}{\xi-x- i \epsilon}-\frac{1}{\xi+x-i \epsilon} \right]
  H (x,\xi,t,{\cal Q}^2)\,,
\label{eq:calH}
\end{equation}
where $\xi = x_B/(2-x_B)$, $x_B$ is Bjorken's scaling variable,
$t$ is the  momentum transfer squared, and
$H$ is the GPD. Formulae such as (\ref{eq:calH}) suggest  to
parametrize GPDs 
at some input scale $\mathcal{Q}^2_0$, to employ QCD evolution equations to
determine the GPDs at the desired scale $\mathcal{Q}^2$, then to make the
momentum fraction convolution to
obtain CFFs that are used to calculate measured cross-sections and asymmetries
via known formulae \cite{Belitsky:2001ns,Belitsky:2010jw}, and, finally, to
confront the result with experimental data by means of the least-squares
method.
Due to the facts that GPDs cannot be fully constrained even by ideal data and that they
depend at the input scale $\mathcal{Q}^2_0$ on three variables, the space
of possible functions, although restricted by GPD constraints, is huge. As a result,
the theoretical uncertainty induced by the choice of the fitting model
is much more serious than in the PDF case, where the model functions
depend at the input scale only on one variable, namely, the longitudinal momentum fraction $x$.
The fact that exclusive data is typically much less precise than inclusive one
exacerbates the situation further.

In this paper we explore an alternative approach, in which
\emph{neural networks} are used in place of specific models.
This essentially eliminates the problem of model dependence
and, as an additional advantage, facilitates a convenient method to propagate
uncertainties from experimental measurements to the final result.
In the context of nucleon structure studies, neural networks have already
been successfully applied to extract PDFs by the
NNPDF group \cite{Ball:2008by,Ball:2010de,Ball:2011mu} and to parametrize
electromagnetic form factors \cite{Graczyk:2010gw}.
Similarly, they have been employed for parametrization of spectral function of
hadronic tau decays \cite{Rojo:2004iq}.
In the case of GPDs, because of the abovementioned reasons we expect that
the advantages of this method should be even more pronounced.

While our long-term goal are global fits of GPDs, the data presently available
allows only a less ambitious analysis, namely the fit of the dominant
CFF $\mathcal{H}$ using neural networks\footnote{Note that neural
network analysis of deeply inelastic
scattering data also started by fitting directly the structure function $F_2$
\cite{Forte:2002fg,DelDebbio:2004qj}}.
This reduces the mathematical complexity of the problem (increased also
by GPD constraints), while still retaining the relevance for
the study of hadron structure.

The imaginary part of the CFF (\ref{eq:calH}) is at leading order
\begin{equation}
\frac{1}{\pi}\Im{\cal H}(x_B=\frac{2\xi}{1+\xi},t,{\cal Q}^2)
\stackrel{\rm LO}{=}
H(x,\xi=x,t,{\cal Q}^2) - H(-x,\xi=x,t,{\cal Q}^2)  \;,
\label{eq:DVCSLO}
\end{equation}
and so, although we consider only extraction of CFFs, $\Im\mathcal{H}$
provides us with direct information about the shape of GPD $H$ on the
cross-over line $\xi=x$. Moreover, a ``dispersion relation"
\cite{Teryaev:2005uj,Kumericki:2007sa,Diehl:2007jb,Kumericki:2008di}
\begin{equation}
\Re{\cal H} (x_B,t,{\cal Q}^2) \stackrel{\rm LO}{=}
  {\rm PV}\int_{0}^1\!dx\, \frac{2x}{\xi^2-x^2}
  \left[H (x,x,t,{\cal Q}^2)-H (-x,x,t,{\cal Q}^2)\right] - {\cal C}(t,{\cal Q}^2)\,,
\label{eq:RecalH}
\end{equation}
can then be used as a sum rule to pin down this GPD  outside of the
kinematically accessible region.
Extraction of CFFs from DVCS data in a largely model-independent way
was also performed  in \cite{Guidal:2008ie,Guidal:2009aa,Guidal:2010ig, Moutarde:2009fg}.
In contrast to the work presented here,
these analyses are local in the sense that they extract values of
CFFs separately at each measured kinematic point.
We emphasize that in Refs.~\cite{Guidal:2008ie,Guidal:2009aa,Guidal:2010ig} the systematic uncertainties, induced by other non-dominant CFFs,
were explored with model-dependent constraints for the four twist-two related CFFs.

The paper is organized as follows.
In Section \ref{sec:nnmethod} we describe salient features of the neural network approach.
In Section \ref{sec:fit} we present fits to a well-defined subset of available DVCS data,
so that we could study
neural network method in a controlled situation, free of subtleties
which surface in many fits involving data coming from a number of different
experiments. In particular, we used HERMES
measurements of two observables, beam spin asymmetry (BSA) and beam
charge asymmetry (BCA) in leptoproduction of a real proton (of which DVCS
is a subprocess) \cite{:2009rj}.
This data belongs to a kinematic region where these two asymmetries
are determined essentially by the imaginary and the real part of the
Compton form factor $\mathcal{H}$ (\ref{eq:calH}), respectively,
and where dependence of $\mathcal{H}$ on $\mathcal{Q}^2$ is weak and,
therefore, can be neglected
for simplicity.  The neural network parameterization of CFF $\mathcal{H}$
is then used to make a prediction for another
observable, namely, the beam charge-spin asymmetry in the kinematic region to be
explored by the COMPASS II experiment \cite{COMPASSII}.
In Section \ref{sec:comparison} we place particular emphasis on the determination of
uncertainties.
To do so we fit a simple model using
the standard least-squares method
and compare the resulting uncertainties with those obtained
by neural networks. Section \ref{sec:conclusion} contains conclusions.

\section{\label{sec:nnmethod} Fitting data with neural networks}

\begin{figure}[th]
\begin{center}
\includegraphics[scale=0.8]{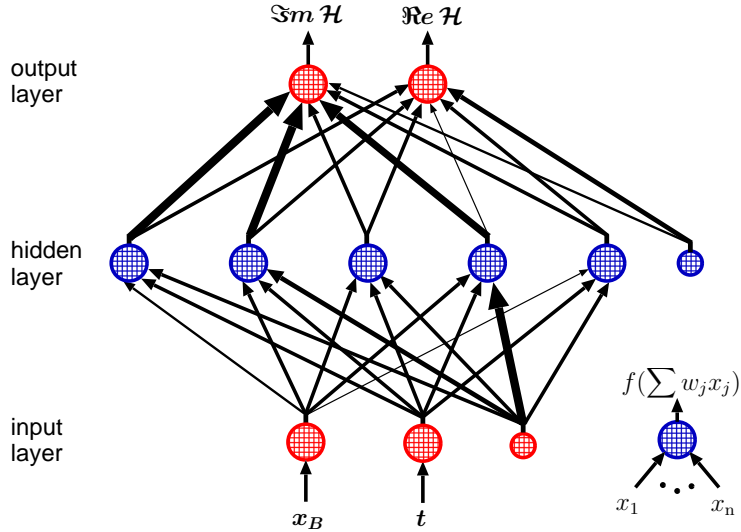}
\end{center}
\caption{A neural network (multilayer perceptron with 2-5-2 architecture) parametrization of
the complex-valued CFF $\mathcal{H}(x_B, t)$. Each blob symbolizes a neuron
and thickness of arrows represents the strengths of weights $w_j$. (Smaller blobs with no
input are so-called \emph{biases} --- they improve the network's properties \cite{Haykin99}.)}
\label{fig:perceptron}
\end{figure}
Neural networks have been applied to a variety of tasks, including optimization problems. One of
the main attractions is their apparent ability to mimic the behavior
of human brain: with their multi-connectedness and parallel processing
of information they can be trained to perform relatively complex
classification and pattern recognition tasks. To this end many kinds
of neural networks and corresponding ``learning'' algorithms
have been developed, see, e.g., \cite{Haykin99}. In our
study, we used one of the most popular neural network types, known
as \emph{multilayer perceptron}.
Schematically shown on \textsc{Fig.}~\ref{fig:perceptron},
it is a mathematical structure consisting of a number of interconnected
``neurons'' organized in several layers.
Each neuron has several inputs and one output. The value
at the output is given as a function
$f(\sum_j w_j x_j)$ of a sum of values at inputs
$x_1, x_2, \cdots$, each weighted by a a certain number
$w_j$. Consequently, a multilayer perceptron is defined by its architecture
(number of layers and number of neurons in each layer),
by the values of its weights, and by the shape of its
\emph{activation function} $f(x)$. Again somewhat inspired by
biological neuron properties, the activation function
is often taken as a nonlinear function with saturating
properties for large and small values of its argument.
We employed a very popular logistic sigmoid function $$f(x) = 1/(1+\exp(-x))$$
for neurons in inner (``hidden'') layer(s), while for input and output layers we
used the identity function\footnote{To increase representational power of
a given network, one can also use sigmoid function in
input and output layers,  but then input and output
values have to be rescaled so that they don't  fall into the
saturation regions of the activation function. For our relatively simple case the identity function is sufficient.} $f(x)=x$.
By iterating over the following steps the network is then trained, i.e., it ``learns'' how to describe
a certain set of data points:
\begin{enumerate}
\item Kinematic values (two in our case: $x_B$ and $t$) of
the first
data point are presented to two input-layer neurons
\item These values are then propagated through the network
according to weights and activation functions.
In the first iteration, weights are set to some random value.
\item As a result, the network produces some resulting values
of output in its output-layer neurons. Here we have
two: $\Im\mathcal{H}$ and $\Re\mathcal{H}$.
Obviously, after the first iteration, these will be
some random functions of the input kinematic values:
$\Im\mathcal{H}(x_B, t)$ and $\Re\mathcal{H}(x_B, t)$.
\item Using these values of CFF(s), the observable(s) corresponding
to the first data point is (are) calculated and it is (they are) compared to
actually measured value(s), with squared error used
for building the standard $\chi^2$ function.
\item The obtained error is then used to modify the network:
It is, possibly weighted by the inverse uncertainty
of the experimental measurement, propagated backwards
through the layers of the network and each weight is adjusted
such that this error is decreased.
The concrete algorithm for this weight adjustment is discussed at the end
of this section.
\item This procedure is then repeated with the next data point, until
the whole training set is exhausted.
\end{enumerate}
This sequence of steps (called training \emph{epoch}) is repeated
until the network is capable to describe experimental data with a sufficient
accuracy --- the precise stopping criterion is specified below.

From what is said it should be clear that a neural
network is nothing more but a complicated non-linear
multi-parameter function and that the training procedure is
equivalent to the least-squares fitting of this function to
data. The actual power of the approach stems from the following:
\begin{itemize}
\item A neural network serves as \emph{unbiased} interpolating
function. Its complicated dependence
on parameters enables it to approximate
any smooth function with comparable ease\footnote{This flexibility of a neural
network is formalized in the \emph{universal approximation
theorem} \cite{Haykin99}. Roughly spoken, this theorem states that any continuous multivariate function
$f(x_1, x_2, \dots)$ can be arbitrary accurately
approximated by a multilayer perceptron. Strictly spoken, one hidden
layer is enough for this, but networks with
more hidden layers can have fewer neurons and/or can be more efficiently trained.}.
\item Updating the weights of a given neuron by back-propagation of error  uses only values and
gradients of activation functions in the immediate network neighborhood. In this sense the training process is local and it
is algorithmically efficient.
\item
This approach enables the following convenient method for propagation of experimental
uncertainties (and even their correlations) into the final result:
In our application  a ``replica data set'' is interpolated. It is obtained from
original data by generating random artificial data points using
Gaussian probability distribution with a width defined by the error bar of experimental measurements.
Taking a large number $N_{rep}$ of such replicas, the resulting family of trained neural networks
$\mathcal{H}^{(1)},\ldots,\mathcal{H}^{(N_{rep})}$
defines a probability distribution of the represented CFF
$\mathcal{H}(x_B, t)$ and of any functional $\mathcal{F}[\mathcal{H}]$
thereof. Thus, the mean value of such a functional and
its variance are \cite{Giele:2001mr,Forte:2002fg}
\begin{align}
 \Big\langle \mathcal{F}[\mathcal{H}] \Big\rangle& =
  \frac{1}{N_{rep}}\sum_{k=1}^{N_{rep}} \mathcal{F}[\mathcal{H}^{(k)}]\;,
\label{eq:funcprob} \\
\Big(\Delta \mathcal{F}[\mathcal{H}]\Big)^2& =
\Big\langle \mathcal{F}[\mathcal{H}]^2 \Big\rangle  -
\Big\langle \mathcal{F}[\mathcal{H}] \Big\rangle^2 \;.
\label{eq:variance}
\end{align}
For details of this ``Monte Carlo'' procedure see Refs.~\cite{Forte:2002fg,Rojo:2006ce}
and note that this is a general method of error propagation, which is
not related to neural networks themselves and can be used
also for error propagation in standard least-squares fitting
of model functions to data.

\end{itemize}

\begin{figure}[t]
\begin{center}
\includegraphics[scale=0.85,clip]{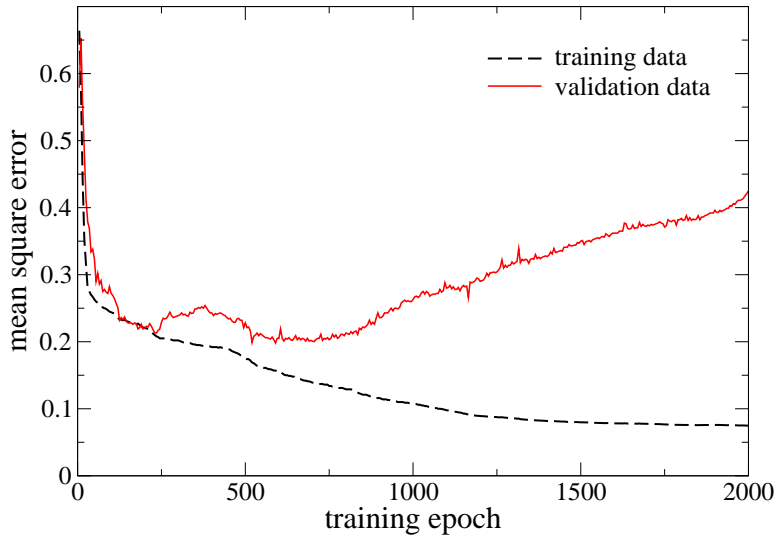}
\end{center}
\caption{Typical progress of neural network training: the error for
the description of training data (dashed) is monotonously decreasing, while
the error for the validation data (solid) increases when the neural network
is overfitted. In the plotted example one would use the
network obtained after 500-700 training epochs as the final
result}
\label{fig:crossvalidation}
\end{figure}

The power of neural networks to approximate any continuous
function implies also the danger of overfitting the data
(also known as \emph{overtraining}). Namely, after a certain number of
iterations the network will not only describe the general
dependence of observables on kinematic variables, but will also adjust to
the random fluctuations of data. This unwanted behavior is
prevented by the cross-validation procedure in which initial
data is divided into two sets: \emph{training} set and
\emph{validation} set. Then performance of the network
is continuously checked on the validation set. Since this
set is not used for training, after the onset of overfitting,
the error of the network's description of validation sets will
increase, see \textsc{Fig.}~\ref{fig:crossvalidation}.
This is the moment at which training is stopped.

For more informations on neural networks we refer the reader to
the vast literature on the subject, e.g. \cite{Haykin99}, while
the PhD thesis \cite{Rojo:2006ce} is a good reference for the
particular approach used also in this work. In the remainder of
this section, we specify some details about the neural networks we used.

In our study we employed the \textsc{PyBrain} software library for
neural networks \cite{pybrainpaper}. This choice was
motivated by (i) the great flexibility of this library, which allows
to explore various types of neural networks,
and (ii) it being written in the Python programming language, such that
it was
easy to link with our already existing Python code for DVCS
analysis.
Like almost every other freely available neural network software
(at least to our knowledge), \textsc{PyBrain} expects that the network output
is directly compared to training data samples for error calculation.
In our case, however, the network output must be transformed, i.e.,
the DVCS
observables must be calculated from $\Im\mathcal{H}$ and $\Re\mathcal{H}$ (see step no. 4
in the training procedure described above), before the error can be evaluated.
For this, we had to make some modifications to the \textsc{PyBrain} software.

Concerning the network architecture, the number of neurons in the
input and output layers were fixed by the problem at hand
to be two, see \textsc{Fig.}~\ref{fig:perceptron}. To fix
the number of hidden layers and their neurons,
one has to consider a trade-off between two requirements: 
the number of neurons in the hidden layer(s) must be large enough
to ensure sufficient expressive power of the network but must not
be too large with respect to the number of available data
points, otherwise the training will become difficult.
In our case, it is already known from previous studies, e.g.
\cite{Kumericki:2009uq}, that CFF $\mathcal{H}$ is a reasonably well-behaved
function. Thus, since we used a relatively small sample of few
tens of data points, see next section, just one hidden layer
with about ten neurons proved to be sufficient.
The specific results presented in this paper were obtained
by training series of neural networks, with 13 neurons
in the hidden layer each.  We checked that
results do not change appreciably if a neuron is added or removed from the
hidden layer, which is one of the standard tests in neural network applications.

There are various training algorithms available for updating the network weights,
from a simple back-propagation algorithm \cite{Haykin99} to
genetic algorithms like the one used by the NNPDF group \cite{Rojo:2006ce,Ball:2010de}.
After some experimentation we ended up using the so called
\emph{resilient back-propagation}
algorithm \cite{rprop-}.
This algorithm is a modification of the usual back-propagation algorithm:
only the signs of the partial derivatives
of the error function, and not their magnitude are taken into account.
Resilient back-propagation is known to be very
reliable, fast and insensitive to parameters of the network and of
the learning process.
More sophisticated algorithms are needed when
correlations are included in the $\chi^2$ error function or in cases
for which the connection between the network output and the experimental
observables is very non-linear. The latter would, e.g. be the case when the
network output would represent GPDs  rather than CFFs, so that
the connection would
involve convolutions with QCD evolution operators and
hard-scattering coefficient functions.
Let us finally note that besides perceptrons, there are attempts to apply
other kinds of neural networks  in nucleon structure studies, such as
self-organizing maps \cite{Honkanen:2008mb}.

\section{\label{sec:fit} Example fit to data}

For the first application, to assess the power of the neural network
approach, we took just two sets of measurements of leptoproduction
of a real photon by scattering leptons off unpolarized protons. One set
consisted of 18 measurements of the first sine harmonic
$A_{LU}^{\sin\phi}$ of the
beam spin asymmetry (BSA)\footnote{Actually,
what was measured is the so-called charge-difference BSA, see \cite{:2009rj}
for definition, but in the leading twist approximation this is equal to
the simple BSA (\ref{eq:bsa})
and we disregard the difference of these two observables in this paper.}
\begin{equation}
BSA
\equiv \frac{ {\rm d}\sigma_{e^\uparrow} - {\rm d}\sigma_{e^\downarrow} }
{ {\rm d}\sigma_{e^\uparrow} + {\rm d}\sigma_{e^\downarrow} }
\sim
A_{LU}^{\sin\phi}\sin\phi \;,
\label{eq:bsa}
\end{equation}
(where $\phi$ is the azimuthal angle in the so-called Trento convention),
while in the other set there were 18 measurements of the first
cosine harmonic $A_{C}^{\cos\phi}$ of the beam charge asymmetry (BCA)
\begin{equation}
BCA
\equiv \frac{ {\rm d}\sigma_{e^+} - {\rm d}\sigma_{e^-} }
{ {\rm d}\sigma_{e^+} + {\rm d}\sigma_{e^-} }
\sim
A_{C}^{\cos 0\phi} + A_{C}^{\cos\phi}\cos\phi \;.
\label{eq:bca}
\end{equation}
Both sets cover identical kinematic regions
$$0.05 < x_B < 0.24\,,\quad
0.02  < -t/{\rm GeV}^2< 0.46\,,\quad \mbox{and}\quad
1.2 < \mathcal{Q}^2/{\rm GeV}^2 < 6.11\,.$$
As mentioned in the Introduction,
we assumed that QCD evolution effects can be neglected, i.e., that CFFs
are independent of $\mathcal{Q}^2$.
Furthermore, it has been shown that the hypothesis of CFF $\mathcal{H}$ dominance
leads to a successful description of this data \cite{Kumericki:2009uq} and so we relied on this
assumption (with the understanding that it
is only an approximation which should be given up, once sufficiently precise data is available).
Thus, at present, just a single CFF $\mathcal{H}(x_B, t)$, or two real-valued
functions $\Im\mathcal{H}(x_B, t)$ and $\Re\mathcal{H}(x_B, t)$, are extracted from
data by our neural networks.

A comment about the relation between $\Im\mathcal{H}$ and $\Re\mathcal{H}$ is in order.
Analytic properties of CFF $\mathcal{H}$ relate these two functions
via a ``dispersion relation'', which reads in twist-two approximation:
\begin{eqnarray}
\Re\mathcal{H}(x'_B=\frac{2\xi}{1+\xi},t) =
{\rm PV} \frac{1}{\pi} \int_{0}^1\!dx\, \frac{2x}{\xi^2-x^2}
 \Im\mathcal{H}(x_B=\frac{2x}{1+x},t)
- {\cal C}(t) \;,
\label{eq:dr}
\end{eqnarray}
cf.~Eqs.~(\ref{eq:DVCSLO}) and (\ref{eq:RecalH}). As discussed in the Introduction,
this could be used to simplify the fitting function set
from $$\{ \Im\mathcal{H}(x_B, t), \Re\mathcal{H}(x_B, t) \}\quad\mbox{to}\quad  \{ \Im\mathcal{H}(x_B, t), \mathcal{C}(t) \}.$$
However, for two reasons we treat here the imaginary and real part of CFF $\mathcal{H}$ as independent quantities.
First, assumption of ``dispersion relation'' (\ref{eq:dr}) by itself introduces a theoretical
bias we want to avoid. In Section \ref{sec:comparison} we will see that
model-dependent least-squares fits that are based on the ``dispersion relation" (\ref{eq:dr})
link, e.g.,  assumptions about the low $x$ behaviour of $\Im\mathcal{H}$ to that of $\Re\mathcal{H}$
which might lead to rather misleading results. Instead, neural networks which do not make use of
(\ref{eq:dr}) give large uncertainties for both, which is probably more realistic.\\
Second, to implement the ``dispersion relation" approach in a neural network framework
one would need to either (i) disconnect topologically the network
output representing $\mathcal{C}(t)$ from the input representing $x_B$,
or (ii) to create a separate neural network with one input and one
output (1--1) for $\mathcal{C}(t)$ and train it together with the
two-inputs-one-output (2--1) network
representing $\Im\mathcal{H}(x_B, t)$.  Certainly, taking in future
all available fixed target data into account,
such a strategy might be feasible, and if so, would allow to get rid of the $\mathcal{H}$ dominance
hypothesis and might thus improve
the phenomenological value of the final result.

\begin{figure}[t]
\begin{center}
\includegraphics[scale=0.5,clip]{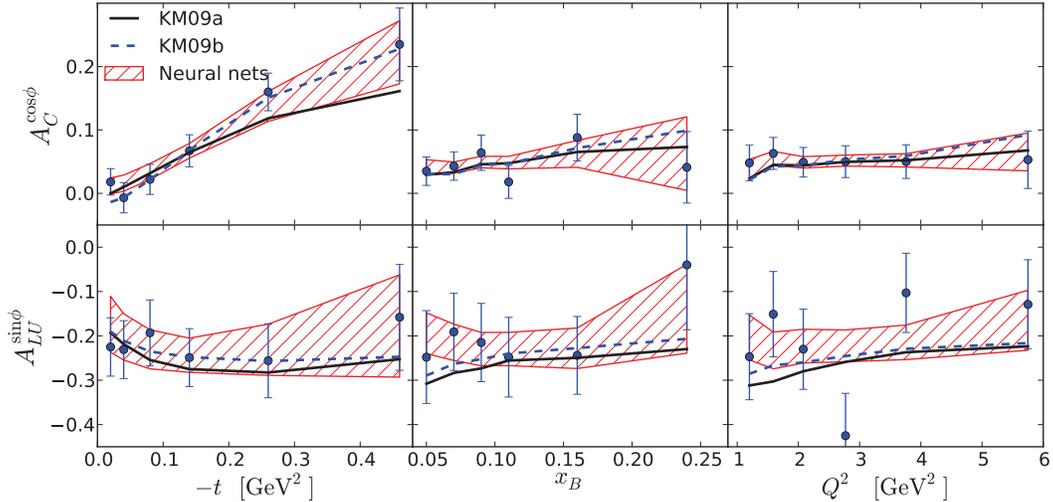}
\end{center}
\caption{First cosine harmonic of beam charge asymmetry $A_{C}^{\cos\phi}$
(\protect\ref{eq:bca}) and first sine
harmonic of beam spin asymmetry $A_{LU}^{\sin\phi}$ (\protect\ref{eq:bsa})
resulting from neural network fit (hatched areas),
shown together with data \cite{:2009rj}, used for training.
Two model fits, KM09a (solid) and KM09b (dashed) from \cite{Kumericki:2009uq}
are also shown for comparison.}
\label{fig:hermes09}
\end{figure}

Using the procedure described in Section~\ref{sec:nnmethod}, we created 50 replicas from
the set of 36 measured data points, which were randomly divided into
a training set of 25 points and a validation set of 11 points.
Then we trained one neural network on each training set,
monitoring progress on the validation set, as in \textsc{Fig.}~\ref{fig:crossvalidation}.
Training was performed either until the onset of overlearning, or
for 2000 epochs, whichever came first.
We ended up with 50 neural networks, each representing CFF $\mathcal{H}$.
Most of the neural networks fit the original data set quite well
(44 of our 50 neural networks have $\chi^2$ less than the number of data points, 36),
with few poorer fits as expected due to the stochastic nature of the set of
data replicas.

Using this set of neural networks as a probability distribution in a functional
space of all $\mathcal{H}$ one can predict any function of
$\mathcal{H}$, together with its uncertainty, using
(\ref{eq:funcprob}) and (\ref{eq:variance}). As a consistency check  we plot in
\textsc{Fig.}~\ref{fig:hermes09}
the result for the very observables that are used for training of
networks.
We observe that uncertainties of measurements have been correctly propagated
into the corresponding uncertainties given by neural networks.
This is particularly visible in the first two panels of \textsc{Fig.}~\ref{fig:hermes09},
where uncertainties of both data and neural networks increase with $|t|$
in the same fashion.

Error bands here and in other figures denote the
uncertainty that corresponds to one standard deviation (one sigma).
We checked for departures from a Gaussian distribution of network
results and found
them to be small in the region where data is available. This means that a one-sigma error band
corresponds indeed to a 68\,\% confidence level. However, as observed also
by the NNPDF group \cite{Ball:2010de}, in the extrapolation region, where
there is no data, departures from Gaussian distribution are significant,
and 68\,\% confidence level regions are generally smaller than the one-sigma
regions we plot.

\begin{figure}[t]
\begin{center}
\includegraphics[scale=0.5,clip]{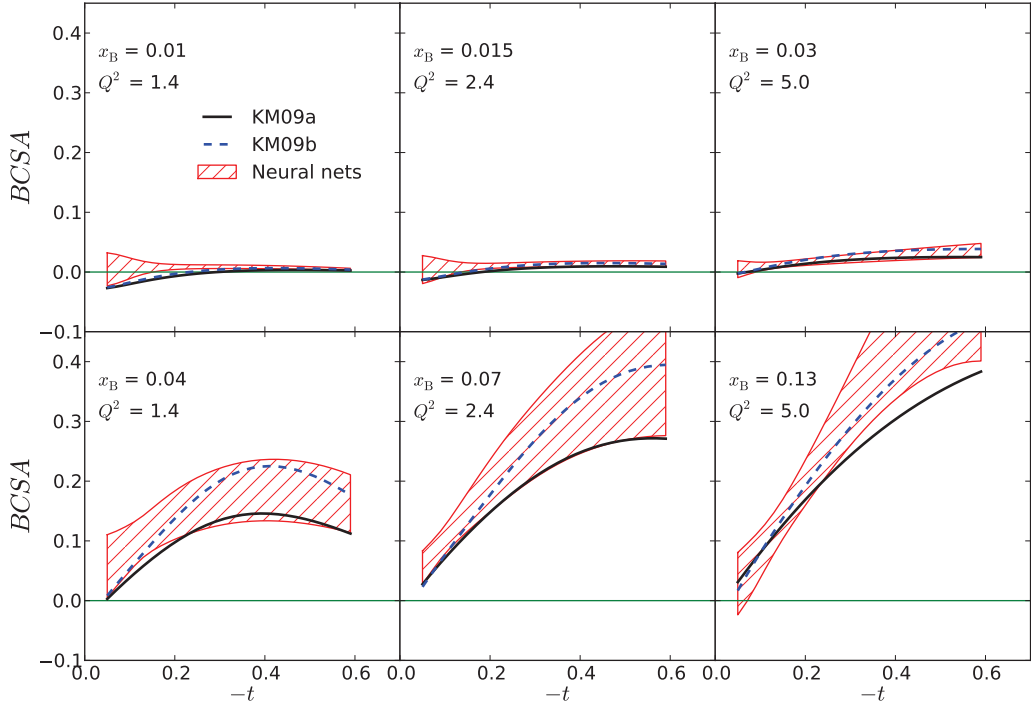}
\end{center}
\caption{Predictions of neural network fits for beam charge-spin asymmetry
(\ref{eq:bcsa}) in COMPASS II kinematics
($E_{\mu} =160\, {\rm GeV}$,  $\phi = 0$)
(hatched areas).
Two model fits, KM09a (solid) and KM09b (dashed) from \cite{Kumericki:2009uq}
are also shown for comparison.}
\label{fig:compass}
\end{figure}

As an example of a proper prediction coming from our analysis
we plot in \textsc{Fig.}~\ref{fig:compass} the beam charge-spin asymmetry (BCSA),
\begin{equation}
BCSA
\equiv \frac{ {\rm d}\sigma_{\mu^{\downarrow +}} - {\rm d}\sigma_{\mu^{\uparrow -}} }
{ {\rm d}\sigma_{\mu^{\downarrow +}} + {\rm d}\sigma_{\mu^{\uparrow -}} }
\label{eq:bcsa}
\end{equation}
as a function of $t$, for several kinematic points that are
characteristic for the COMPASS II experiment (where the muon is taken to
be massless and the polarization is set equal to 0.8).
This experiment was chosen because its kinematics overlaps with that of the
HERMES data used for neural network training. Hence these predictions
represent partly interpolation and partly extrapolation of HERMES data,
thus testing this framework in a nontrivial way.

\section{\label{sec:comparison} Comparison with model fits and assessment of uncertainties}

One of the main advantages of the neural network approach to hadron structure
is that it offers a convenient way to assess the uncertainties of non-perturbative
functions (GPDs, CFFs, PDFs, \dots). Let us now emphasize this point further by making a
comparison with the common method  in which one chooses some model
and makes a least-squares fit of it to the data.
To quantify this comparison, we took a simple model of the CFF $\mathcal{H}$
and fitted it to the very same data to which neural networks were trained.

We adopted a version of the model used in \cite{Kumericki:2009uq}
for which the partonic decomposition of $\Im\mathcal{H}$ is:
\begin{eqnarray}
\label{SpeFunH-ParDec}
\Im{\cal H}(x_{\rm Bj},t) =
\pi \left[
\left(2\,\frac{4}{9} + \frac{1}{9}\right)
H^{\rm val}(\xi,\xi,t)
+ \frac{2}{9} H^{\rm sea}(\xi,\xi,t)
\right]\,.
\end{eqnarray}
We parametrized the GPDs along the cross-over trajectory $\xi=x$ as:
\begin{eqnarray}
\label{GPD-Ans}
H(x,x,t)  =
\frac{n\, r}{1+x}
\left(\frac{2 x}{1+x}\right)^{-\alpha(t)}
\left(\frac{1-x}{1+x}\right)^{b}
\frac{1}{\left(1-  \frac{1-x}{1+x} \frac{t}{M^2}\right)^{p}}\,~~~.
\end{eqnarray}
Here $2^{-\alpha(0)}\, n$ is the normalization of PDF $q(x)=H(x,0,0)$ from PDF fits,
$r$ is the skewness ratio at small $x$ (the ratio of a GPD at
some point on the
cross-over trajectory and the corresponding PDF),
$\alpha(t)$ is the ``Regge trajectory'',
$b$ controls the  large-$x$ behavior,
and $M$ and $p$ control the $t$-dependence \cite{Hwang:2007tb}.
The parameters of the sea-quark GPD  $H^{\rm sea}$ were
taken to be as in \cite{Kumericki:2009uq}
($\alpha^{\rm sea}(t) = 1.13 + 0.15\, t/{\rm GeV}^2$,
$n^{\rm sea}=1.35$, $r^{\rm sea} = 1$, $b^{\rm sea} = 2$,
$(M^{\rm sea})^2 = 0.5\, {\rm GeV}^2$, $p^{\rm sea}=2$).
For the valence quark GPD $H^{\rm val}$, we also
fixed  $\alpha^{\rm val}(t) =  0.43 + 0.85\, t/{\rm GeV}^2$,
$n^{\rm val} =1.35$, and $p^{\rm val}=1$.
This left $r^{\rm val}$, $b^{\rm val}$ and $M^{\rm val}$
as free parameters, to be determined by the fit to the data.
We obtained $\Re\mathcal{H}$ from the ``dispersion
relation" (\ref{eq:dr}), treating the subtraction
constant $C$ as a fourth and final free parameter
of the model\footnote{We assume the subtraction constant
in (\ref{eq:dr})  to be $t$-independent, $\mathcal{C}(t) = C$.
This is motivated a posteriori by the fact that such a simple
model is good enough to fit the data. Additionally, when
we parametrize $\mathcal{C}(t) = 1/(1-t/M_{C}^2)^2$,
the parameter $M_{C}$ tends to be strongly correlated with
other parameters of the model, making error analysis,
which is our main task, more difficult.}.

Fitting this model to the same 36 data points used for
neural network fit, using the well-known \textsc{Minuit} software
package \cite{James:1975dr} for minimization of the $\chi^2$ function,
resulted in a good fit ($\chi_{0}^2/{\rm n.d.o.f} = 22.2/32$) with
parameter values
\begin{equation}
   r^{\rm val} = 1.11\,, \quad
   b^{\rm val} = 1.79\,, \quad
   M^{\rm val} = 0.51\, {\rm GeV}\,, \quad
   C = 2.25 ~~~~.
\label{eq:tDR3}
\end{equation}
The uncertainty of the resulting fit is commonly determined by the
so-called Hessian method. The Hessian
matrix is given by the second derivatives of $\chi^2$ with respect to
the model parameters $a_i$ at the minimum $\chi = \chi_{0}$:
\begin{equation}
H_{ij} =  \left. \frac{\partial^2 \chi^2}{\partial a_i \partial a_j}\right|_{\chi=\chi_0} \;.
\label{eq:hesse}
\end{equation}
The uncertainty $\Delta f$ of the function $f(a_i)$ of
these parameters,
including the uncertainty of the parameters themselves, is given
by the error propagation formula
\begin{equation}
(\Delta f)^2 = \Delta\chi^{2} \sum_{ij} \frac{\partial f}{\partial a_i}
H^{-1}_{ij} \frac{\partial f}{\partial a_j} \;.
\label{eq:errorprop}
\end{equation}
In a textbook approach to statistics, one sigma uncertainty is obtained
by taking $\Delta\chi^2 = 1$ in this formula.
However, there are several problems with this simple procedure.
First, $\chi^2$ tends to vary very differently in different
directions of parameter space, yielding a Hessian with large variations in the eigenvalue spectrum
which might then imply numerical difficulties. In our case, where eigenvalues vary over three orders
of magnitude, the problem was not so severe. However, in DVCS fits, involving more parameters,
we have also observed much stronger variations. To improve the reliability of
the Hessian method, we followed an iterative procedure \cite{Pumplin:2000vx}
which allows to find natural directions in parameter space and natural step sizes for
the finite-difference calculation of the Hessian matrix.
Second, it is known that with global fits, i.e., when one combines data from several
different experiments, errors are not distributed in the expected way.
Often, reasoning strictly within textbook
statistics, one would have to reject some data as incompatible
either with itself (i.e., because its $\chi^2$ is too large) or with
other data sets (i.e., because likelihood curves poorly overlap).
This is mainly due to systematic errors which are not understood.
Without entering further into the discussion of this complex issue,
we can follow the CTEQ procedure \cite{Pumplin:2002vw}, where $\Delta \chi^2$
in (\ref{eq:errorprop}) is increased ad hoc to $\Delta \chi^2 = T^2$
in order to accommodate all experimental measurements used for fitting.
The tolerance parameter $T$ is determined by calculating the average distance from the
best fit along eigenvectors of the Hessian matrix at which the
model is still compatible with some particular experiment at some
confidence level (C.L.). In global PDF fits, with many data sets,
values of $T \sim 5-10$ are common, see e.g. \cite{DeRoeck:2011na}
for a recent review. Nevertheless,
using this procedure, we arrived at a tolerance $T\approx 1$
(for 68\,\% C.L.; CTEQ use 90\,\%). We conclude that in our case with just
two sets of data, coming from the same experiment, there are no
statistical inconsistencies and we can actually use the textbook formula
(\ref{eq:errorprop}) with $\Delta \chi^2 =1$, which we did\footnote{Preliminary
analysis taking all available data for DVCS on
unpolarized target suggests that the tolerance $T$ will have to be
increased for a true global fit.}.
We note in passing that we found that model parameters are significantly more
constrained by the BCA than by the BSA data.

\begin{figure}[t]
\centerline{\includegraphics[scale=0.5,clip]{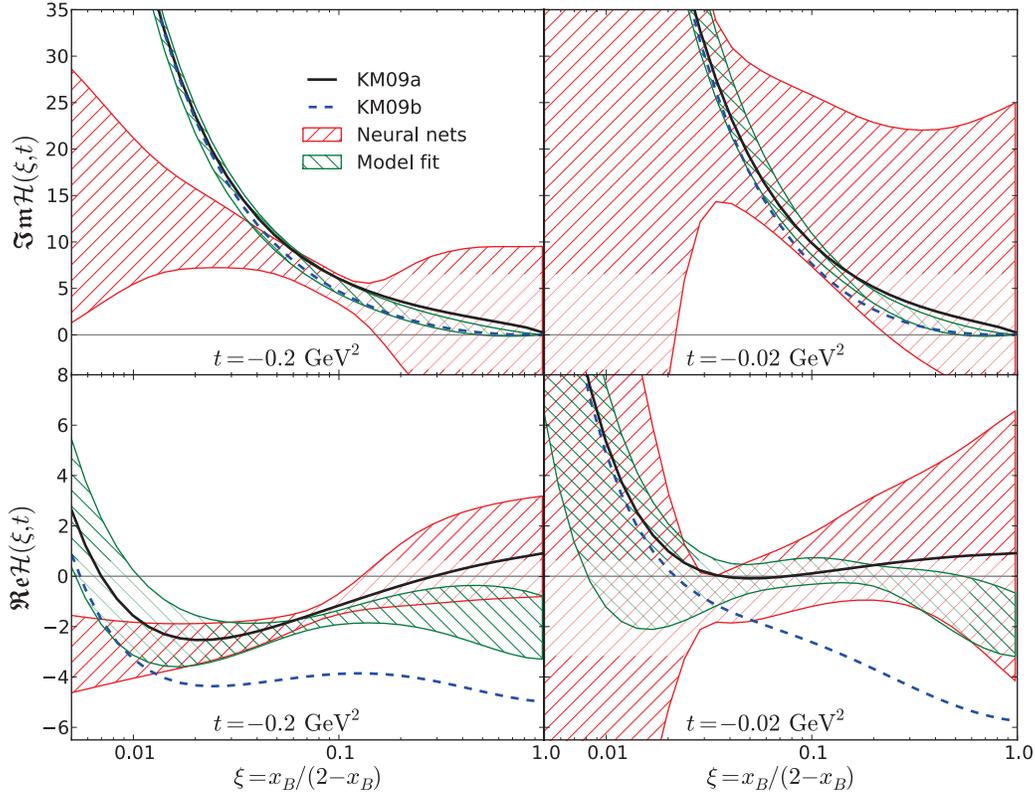}}%
\caption{\small
Neural network extraction of $\Im{\cal H}(x_{\rm Bj},t)$
and $\Re{\cal H}(x_{\rm Bj},t)$ (ascending hatches)
from HERMES BCA and BSA~\cite{:2009rj} data compared with three model
fits, one of which with determined uncertainties (descending hatches).
\label{fig:cff} }
\end{figure}

We can now compare the model fit (\ref{eq:tDR3}), taking also into account uncertainties calculated
by the procedure just described, with the neural network parameterization of the CFF $\mathcal{H}$.
Both findings are displayed in \textsc{Fig.}~\ref{fig:cff}, together with
two models from \cite{Kumericki:2009uq}, where
$\Im\mathcal{H}$ (upper panels) and $\Re\mathcal{H}$ (lower panels) are separately plotted.
We notice several interesting features:
\begin{itemize}

\item In the kinematic region of measured data (this is roughly the
middle vertical third of the left panels), there is a good agreement of
values and uncertainties of neural networks (ascending hatches)
and model fit (descending hatches). This shows that both fitting
methods correctly \emph{interpolate} the data, and that two, quite
different, statistical methods used for determination of uncertainties
are mutually consistent.

\item As one starts to \emph{extrapolate} the fitted CFF $\cal H$ outside of the
data region (left and right thirds of left panels
and the whole of the right panels), the two methods predict markedly different
shapes and uncertainties for the CFF $\cal H$.
All model fits show effects of theoretical bias by following the
$x^{-\alpha}$ functional behavior at small $x$ and claiming
a small uncertainty there, whereas in neural network parameterizations it is obvious
that extrapolated functions are very unconstrained.
This is particularly visible in the right panels, illustrating
the difficulty of a model-independent extrapolation
towards $t= 0$, which is a limit of particular
interest for hadron structure studies.

\item The model fit and the corresponding uncertainties (descending hatches) show the
effect of the dispersion relation integral constraint (\ref{eq:dr})
 --- $\Im\mathcal{H}$ is significantly constrained also outside
of the data region.
This is actually a welcome feature of the dispersion
relation fits, offering the opportunity to access nucleon structure in
regions not directly accessible in the experiment.
However, its usefulness depends on
the extent to which the functional forms used are firmly established.

\item The model KM09b (dashed line), which was obtained in \cite{Kumericki:2009uq}
by including additional data, and, more importantly,
by extending the model to include contributions of the CFFs
$\widetilde{\mathcal{H}}$ and $\widetilde{\mathcal{E}}$, disagrees
for $\Re\mathcal{H}$ with the other three fits.
This emphasizes that the assumed $\mathcal{H}$ dominance
might substantially affect the quality of our results.

\end{itemize}

\section{\label{sec:conclusion} Conclusions and outlook}

Based on HERMES  measurements of lepton scattering off unpolarized protons,
a reaction which should be dominated by the CFF ${\cal H}$,
we performed the first neural network analysis of deeply virtual Compton
scattering data and obtained a neural network representation of
the CFF $\mathcal{H}$. This was then used to make a prediction
for the beam charge-spin asymmetry, measurable at COMPASS II.
The Monte Carlo method of propagation of experimental errors
enabled us to determine also the uncertainty of the
resulting CFF, and we have
found that (in the kinematic region where data is available)
it agrees well with
the uncertainty determined by model fits and standard statistical
procedures. However, neural networks combined with
Monte Carlo error propagation are a conceptually cleaner and
practically simpler method. The propagation of uncertainty from
experimental data is intrinsic. Most importantly, this provides
essentially a model-independent
approach to determine  CFFs and GPDs.

One could argue that the neural network approach,
just because it is essentially model-independent, does nothing more than
faithfully representing the information available in the experimental data
by universal objects: CFFs (as presented here) or GPDs (yet to
be extracted by this approach). This is valuable in itself
but can also be considered as a stepping stone towards
improved
model-dependent studies, that in principle offer the advantage
that our detailed theoretical and phenomenological understanding of QCD dynamics
can be taken into account. Such a two-step procedure would be quite natural
as models can be more directly compared to neural-network determined CFFs or GPDs
than to actual measured observables.

Finally, let us stress that we consider the presented neural network study only as a first step in
the analysis of DVCS data. The problem here is that the number of observables measured at
a given kinematic point is in practice not sufficient to pin down all twelve CFFs,
some of which
are kinematically suppressed.


\subsection*{Acknowledgments}

K.K. thanks Juan Rojo for many valuable comments and Zvonimir Vlah for
discussions on neural networks.
D.M. and K.K. are grateful to the Institut f\"{u}r Theoretische Physik, University
of Regensburg and to
Theory Group of the Nuclear Science Division at Lawrence Berkeley National Laboratory
for their warm hospitality.
This work was supported by the BMBF grant under the contract no. 06RY9191,
by EU FP7 grant HadronPhysics2, by DFG grant, contract no. 436 KRO 113/11/0-1 and by
Croatian Ministry of Science, Education and Sport, contract no.
119-0982930-1016.


\providecommand{\href}[2]{#2}\begingroup\raggedright\endgroup

\end{document}